\documentclass[%
 reprint,
superscriptaddress,
 amsmath,amssymb,
 aps,
 prl
]{revtex4-2}

\usepackage{graphicx}
\usepackage{dcolumn}
\usepackage{bm}
\usepackage[T1]{fontenc}
\usepackage{braket}
\usepackage[utf8x]{inputenc}
\usepackage{easyReview}
\usepackage{soul}

\begin{document}

\preprint{APS/123-QED}

\title{Control of dimer chain topology by Rashba-Dresselhaus spin-orbit coupling}


\author{Pavel~Kokhanchik}
\affiliation{Institut Pascal, Université Clermont Auvergne, CNRS, Clermont INP, F-63000 Clermont–Ferrand, France}

\author{Dmitry~Solnyshkov}
\affiliation{Institut Pascal, Université Clermont Auvergne, CNRS, Clermont INP, F-63000 Clermont–Ferrand, France}
\affiliation{Institut Universitaire de France (IUF), 75231 Paris, France}
 
\author{Thilo~Stöferle}
\affiliation{IBM Research Europe − Zurich, CH-8803 Rüschlikon, Switzerland}

\author{Barbara~Pi\k{e}tka}
\affiliation{Institute of Experimental Physics, Faculty of Physics, University of Warsaw, Pasteura 5, 02-093 Warsaw, Poland}

\author{Jacek~Szczytko}
\affiliation{Institute of Experimental Physics, Faculty of Physics, University of Warsaw, Pasteura 5, 02-093 Warsaw, Poland} 

\author{Guillaume~Malpuech}
\affiliation{Institut Pascal, Université Clermont Auvergne, CNRS, Clermont INP, F-63000 Clermont–Ferrand, France}

\date{\today}

\begin{abstract}
We study theoretically a dimer chain in the presence of Rashba-Dresselhaus spin-orbit coupling (RDSOC) with equal strength. We show that the RDSOC can be described as a synthetic gauge field that controls not only the magnitude but also the sign of tunneling coefficients between sites. This allows to emulate not only a Su-Schrieffer-Heeger chain which is commonly implemented in various platforms, but also, all energy spectra of the transverse field Ising model with both ferromagnetic and antiferromagnetic coupling. We simulate a realistic implementation of these effective Hamiltonians based on liquid crystal microcavities. In that case, the RDSOC can be switched on and off by an applied voltage, which controls the band topology, the existence and characteristics of topological edge states, or the nature of the ground state. This setting is promising for topological photonics applications and from a quantum simulation perspective.
\end{abstract}

\maketitle

Artificial gauge fields have been synthesized in various platforms for more than a decade \cite{aidelsburger2018artificial}. They constitute the basis of synthetic topological matter \cite{ozawa2019topological2} and quantum simulations \cite{bloch2012quantum}. The key ingredients allowing for topological Hamiltonians are spin-orbit coupling (SOC) and the spatial structuring of potentials for particles (lattices). In photonics, the vectorial (transverse) nature of light modes is responsible for the existence of an intrinsic SOC~\cite{bliokh2004modified,onoda2004hall,kavokin2005optical,leyder2007observation,hosten2008observation,bliokh2008geometrodynamics,bliokh2015spin}. Any spatial inhomogeneity of the refractive index removes the degeneracy between these transverse modes, coupling the light momentum and its polarization pseudospin~\cite{Zhang2020}. In planar cavities hosting 2D massive photonic modes~\cite{kavokin2017microcavities}, this intrinsic SOC is characterized by a winding number 2 around the high symmetry point $k=0$~\cite{kavokin2005optical}.
The combination of photonic SOC with broken time-reversal symmetry (TRS) achieved through the Faraday effect (effective Zeeman splitting) makes 2D photonic bands topologically nontrivial~\cite{shelykh2009proposal,bardyn2015topological,Silveir2015,cras2016,gianfrate2020measurement} characterized by Chern numbers $\pm2$. Topological gaps and related unidirectional edge modes appear when these bulk topological modes are placed in an appropriate photonic lattice~\cite{haldane2008possible,wang2008reflection,wang2009observation,lu2014topological,nalitov2015polariton,karzig2015topological,klembt2018exciton}. The described effect simulates the Quantum Anomalous Hall Effect~\cite{haldane1988model} being the foundation of topological photonics. The field then considerably enriched with the proposals and realizations of quantum pseudospin Hall effect which does not require TRS breaking. In such cases the pseudospin involved can be the angular momentum of ring resonator~\cite{hafezi2011robust,hafezi2013imaging}, polarization ~\cite{khanikaev2013photonic,cheng2016robust}, and sublattice (e.g.~valley) pseudospins, both in 2D~\cite{wu2015scheme,ma2015guiding,barik2016two} and 1D, essentially in that case through the use of dimer chains described by a Su-Schrieffer-Heeger (SSH) Hamiltonian~\cite{su1979solitons,asboth2016short,Solnyshkov2016,St-Jean2017,harder2021coherent}. 

The SSH Hamiltonian is the simplest topological model. It can be easily implemented and supports topological edge modes. Its spectrum is equivalent to the one of the transverse-field Ising model (TFIM), which is the fundamental quantum many-body model describing the transition between ordered (ferro- or antiferromagnetic) and disordered (paramagnetic) phases~\cite{de1963collective,pfeuty1970one,kadowaki1998quantum,calabrese2011quantum,heyl2013dynamical}. However, the sign of tunnelings in SSH realizations is normally positive, which allows to access only the ferromagnetic phase of the TFIM.

A new type of photonic SOC being a superposition of Rashba~\cite{bychkov1984properties} and Dresselhaus~\cite{dresselhaus1955spin} SOCs with equal strength has been recently demonstrated. It was 
initially studied in solid state physics~\cite{bernevig2006exact,koralek2009emergence}, and realized for cold atoms~\cite{lin2011spin}. In photonics, it was implemented first in 1D settings such as metasurfaces with broken inversion symmetry~\cite{dahan2010geometric,frischwasser2011rashba,shitrit2013spin-controlled,shitrit2013spin-optical}, chiral photonic crystal~\cite{yannopapas2011photonic}, photonic lattices of waveguide arrays~\cite{plotnik2016analogue}, asymmetric polariton waveguide~\cite{shelykh2018optical}. Most recently its realization has been reported in planar microcavities either filled with liquid crystal (LC)~\cite{rechcinska2019engineering} or with a birefringent organic crystal~\cite{ren2021nontrivial} where it was called emergent optical activity.
Together with TE-TM SOC, effective Zeeman field and in-plane potential (lattices), RDSOC plays the role of a supplementary control knob for implementing effective Hamiltonians of interest and for controlling the mode topology~\cite{polimeno2021tuning}. 
The crucial property offered by LC microcavities~\cite{yan2014topological,bahari2016zeeman,yao2017theory,liu2021topological,liu2021adiabatic} is the tunability of the modes by means of the external voltage applied to the LC molecules. This allows to switch the RDSOC on and off electrically. So far, the effect of the RDSOC (which in 1D is equivalent to the Rashba SOC) in the SSH setting has been theoretically discussed in the context of electronic systems~\cite{yan2014topological,bahari2016zeeman,yao2017theory,liu2021topological,liu2021adiabatic}, but only for a homogeneous or small Rashba SOC giving no topological effects.

In this Letter, we theoretically study a dimer chain with two polarization eigenstates per site. We include staggered RDSOC and linear birefringence which couple the sublattice and polarization degrees of freedom. We show that RDSOC acts as a synthetic gauge field controlling the magnitude and sign of the tunneling coefficients in a tight-binding approach. The effective Hamiltonian describing this dimer chain realizes a complete mapping of the TFIM spectrum, including antiferromagnetic coupling, not accessible with a standard SSH chain.
We demonstrate a realistic implementation of this chain by using 2D continuous simulations of a zigzag arrangement of sites in a LC microcavity. Here, the staggered RDSOC is controlled by external voltage allowing to switch on and off the band topology and the edge states.  

\begin{figure}[tbp]
\includegraphics[width=0.95\columnwidth]{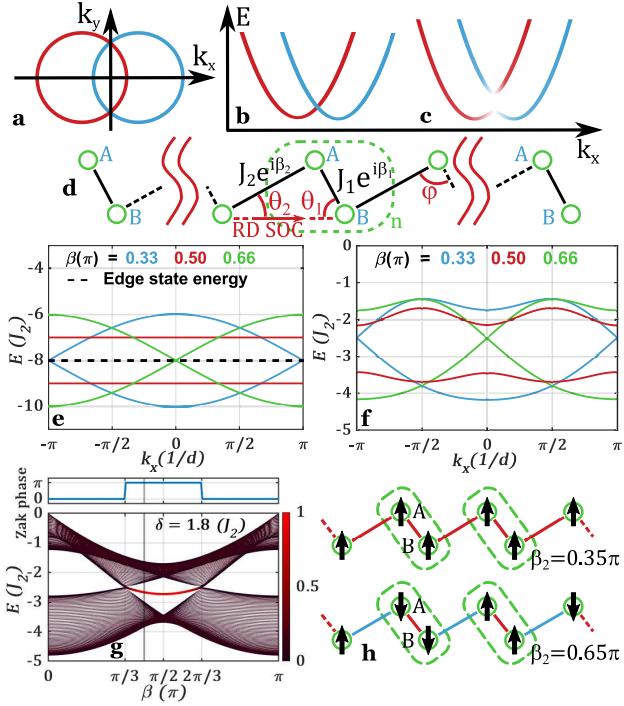}
\caption{Two eigenmodes of $\hat{H}_{LC}$ in reciprocal space for $E=const$ (a) and along $k_x$ for $\delta=0$ (b) and $\delta \neq 0$ (c) (color shows circular polarization); (d) the scheme of zigzag SSH chain with RDSOC; the dispersion of reduced Hamiltonian (e) (Eq.~\eqref{reduced_SSH_Hamiltonian},$\delta=8$) and general SSH Hamiltonian (f) (Eq.~\eqref{SSH_Hamiltonian_reciprocal_space}, $\delta=1.8$) with staggered RDSOC for several values of $\beta$ (for all $\beta$ see \cite{suppl}); (g) energy spectrum of finite SSH chain and Zak phase of the corresponding bulk dispersion lowest band depending on phase $\beta$ produced by RDSOC (the colorscale shows the edge localization); (h) SSH chain corresponding to ferro- (top) and antiferromagnetic  (bottom) phases of TFIM: black arrows correspond to ground state wavefunction phase; red (blue) shows positive (negative) couplings. \label{fig_1}}
\end{figure}

A generic Hamiltonian describing 2D massive particles including RDSOC reads:
\begin{equation}
\label{LC_Hamiltonian}
\hat{H}_{LC} = \frac{\hbar^2 \mathbf{k}^2}{2m} - 2\alpha k_x \hat{\sigma_z} + \delta \hat{\sigma_x},
\end{equation}
where $m$ is the mass, $\mathbf{k}=(k_x,k_y) $, $k_x$ and $k_y$ are the wavevector components, $-2\alpha k_x \hat{\sigma_z}$ is the RDSOC with magnitude $\alpha$, $\delta$ is the splitting between linearly polarized resonant modes, $\hat{\sigma_x}$ ($\hat{\sigma_z}$) is the first (third) Pauli matrice acting on the light polarization pseudospin. This Hamiltonian describes planar microcavity with large linear birefringence \cite{rechcinska2019engineering, ren2021nontrivial}. It provides a high-scale splitting bringing in resonance two modes of different parity (typically $N+1$ and $N$, where $N$ is the quantization number of the longitudinal Fabry-Perot mode) which creates RDSOC. $\delta = (E_{Y,N+1} - E_{X,N})/2$ is a low-scale remaining polarization splitting. 

\begin{figure}[tbp]
\includegraphics[width=0.95\columnwidth]{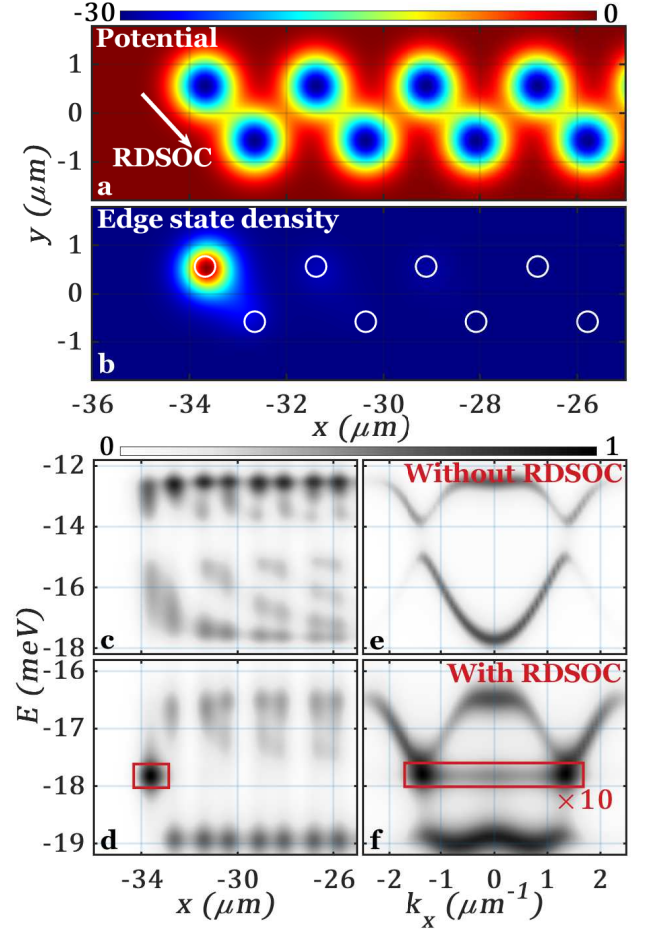}
\caption{(a) 2D SSH potential (in meV) constructed out of Gaussian wells with RDSOC aligned along intracell coupling; (b) normalized edge state density; energy spectrum in real (c,d) and reciprocal (e,f) spaces in the absence (c,e) and presence (d,f) of RDSOC; white-black shows normalized density; the red rectangular shows the edge state (in (f) intensity is increased by 10x for visibility).\label{fig_2}}
\end{figure}

The RDSOC acts as an effective Zeeman splitting proportional to one projection of the wavevector, like optical activity  \cite{ren2021nontrivial}. It is directed along $k_x$ for Eq.~\eqref{LC_Hamiltonian}, as seen  from the cross-section  $k_y = 0$ of the energy bands shown in Fig.~\ref{fig_1}(b) and~\ref{fig_1}(c) for $\delta = 0$ and $\delta \neq 0$, respectively. Along the $k_y$ axis the splitting is absent and the modes are degenerate (Fig.~\ref{fig_1}(a)).

We now consider a tight-binding 1D lattice model including RDSOC. If the links are aligned along the RDSOC direction, the two circularly polarized photons (eigenmodes of $\hat{\sigma_z}$) acquire opposite Aharonov-Casher phase $\pm \beta$ while propagating along the links~\cite{aharonov1984topological,liu2021topological}):  
\begin{equation}
 \beta = \frac{\alpha d}{\hbar^2 / 2m},
\end{equation}
with the link length $d$, RDSOC parameter $\alpha$, and the cavity photon mass $m$. In the circular polarization basis, the tunneling coefficient reads $Je^{i\beta\hat{\sigma_z}}$ which can be understood as the action of a spin-dependent synthetic gauge potential. 
If the link creates an angle $\theta$ with the RDSOC, this leads to a tunneling coefficient $Je^{i\beta \cos \theta \hat{\sigma_z}}$. In a uniform coupling configuration, the change of $\beta$ or $\theta$ does not modify the topology of the chain eigenstates~\cite{suppl}.

The next stage is to dimerize the chain considering both staggered tunellings $J_1$ and $J_2$ and RDSOC orientation. This can be done by making a zigzag with a given RDSOC orientation (Fig.~\ref{fig_1}(d)). The unit cell consists of two lattice sites $A$ and $B$, the lattice constant is $d$. The angles $\theta_1$ and $\theta_2$ between the links and the RDSOC direction control the phases $\beta_1 = \beta \cos \theta_1$ and $\beta_2 = \beta \cos \theta_2$. The RDSOC couples the sublattice pseudospin ($A$ and $B$) and photon polarization pseudospin ($+$ and $-$). The system is more complex than a spinless SSH chain and is described by a $4\times 4$ Hamiltonian which in the basis $(A_+\,A_-\,B_+\,B_-)^T$ reads:

\begin{widetext}
\begin{equation}
\label{SSH_Hamiltonian_reciprocal_space}
\hat{H}_{SSH}(k_x) = -
\begin{pmatrix}
0 & \delta & J_1 e^{i\beta_1} + J_2 e^{-i\beta_2} e^{-i k_x d} & 0 \\
\delta & 0 & 0 & J_1 e^{-i\beta_1} + J_2 e^{i\beta_2} e^{-i k_x d} \\
J_1 e^{-i\beta_1} + J_2 e^{i\beta_2} e^{i k_x d} & 0 & 0 & \delta \\
0 & J_1 e^{i\beta_1} + J_2 e^{-i\beta_2} e^{i k_x d} & \delta & 0
\end{pmatrix}.
\end{equation}
\end{widetext}
The eigenvalues are:
\label{SSH_dispersion}
\begin{equation}
\begin{split}
E(k_x) = & \pm \left[ \delta^2 + \rho_1^2 + \rho_2^2  \right.\\ 
& \left. \pm 2 \rho_1 \sqrt{\delta^2 + \rho_2^2 \sin^2 (\varphi_2-\varphi_1)} \right] ^{1/2},
\end{split}
\label{SSH_energies}
\end{equation}
where $\rho_1 e^{i \varphi_1}=J_1 \cos \beta_1 + J_2 \cos \beta_2 \cdot e^{-i k_x d}$ and $\rho_2 e^{i \varphi_2}=J_1 \sin \beta_1 - J_2 \sin \beta_2 \cdot e^{-i k_x d}$ describe the co- and cross-polarized tunnelings in the linear polarization basis \cite{suppl}. The RDSOC allows to modulate the intra-, intercell co- and cross-polarized tunnelings via $\cos\beta_i$ and $\sin\beta_i$.
The four bands are separated in two pairs, symmetric with respect to $E=0$ because of the global chiral symmetry of the model. If 
$\delta \gg J_1,J_2$, each doublet can be described by a standard SSH Hamiltonian which for the lower H-polarized doublet reads:  
\begin{equation}
\label{reduced_SSH_Hamiltonian}
\hat{H}_{lower}(k_x) \approx - 
\begin{pmatrix}
\delta & J_1 \cos \beta_1 + J_2 \cos \beta_2 e^{-i k_x d}    \\
c.c.   & \delta
\end{pmatrix},
\end{equation}
where $c.c.$ stands for complex conjugate. 
Thus, modulated RDSOC reduces the intracell (intercell) tunneling by a factor $\cos\beta_1 \ (\cos \beta_2)$. It allows to transform a monomer chain into a dimer chain or to swap relations between the links changing the chain topology, characterized by a topological invariant called Zak phase \cite{zak1989berry,suppl}. The Zak phase value (0 or $\pi$) determines the absence or presence of edge states. In spinless SSH, edge states exist if $J_1-J_2<0$ (chain ending with a weak link). With RDSOC, the condition for the edge states existence reads:
\begin{equation}
\label{Topological_state}
\begin{cases}
\rho_1 e^{i \varphi_1}(k_x = \pi/d) = J_1 \cos \beta_1 - J_2 \cos \beta_2 < 0, \\
\rho_1 e^{i \varphi_1}(k_x = 0) = J_1 \cos \beta_1 + J_2 \cos \beta_2 > 0. 
\end{cases}
\end{equation}

The first condition is associated with the standard gap closure for a SSH chain at $k_x=\pi/d$. It can be understood as coming from the renormalization of the tunelling coefficients. The second condition is linked to the new possibility offered by this setting to close the gap at $k_x=0$. This illustrates that the RDSOC allows to change the sign of the tunneling coefficients. Even if a chain is topologically trivial, the same chain with modulated RDSOC could be either nontrivial or trivial, depending on the values of $\beta_1$ and $\beta_2$. Interestingly, conditions~\eqref{Topological_state} do not depend on $\delta$.

\begin{figure*}[tbp]
\includegraphics[width=170mm]{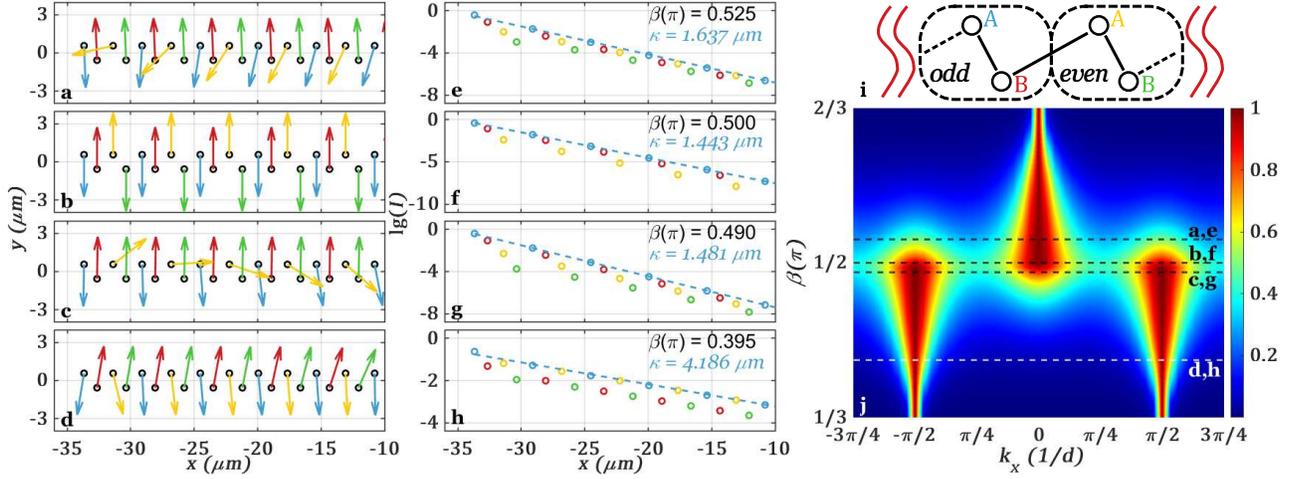}
\caption{The Stokes vector (a-d) and density decay (e-h) in real space of finite SSH chain for different magnitudes of RDSOC $\beta$; $\kappa$ is an exponential decay length obtained by fitting the data (dashed lines); (i) a separation of unit cells into two types; the colors of letters correspond with the colors of arrows in (a-d) and dots in (e-h); (j) the edge state density (color) in reciprocal space depending on RDSOC magnitude $\beta$.\label{fig_3}}
\end{figure*}

Fig.~\ref{fig_1}(e) shows the energy bands computed in the large $\delta$ limit for $J_1=2,J_2=1,\beta_2=0$, and for different values of $\beta_1$ (from now called $\beta$). The critical $\beta$ values for which the band topology is changing are $\pi/3$ and $2\pi/3$ which corresponds to a gap closing at $k_x=\pm\pi/d$ and $k_x=0$ respectively. In between these $\beta$ values, a topological band gap is opened. The edge states lie in the middle of this gap because the reduced $2\times2$ Hamiltonian \eqref{reduced_SSH_Hamiltonian} preserves the chiral symmetry. Precisely at $\beta=\pi/2$, the interference induced by the RDSOC completely suppresses the "strong" link co-polarized tunneling. The corresponding band gap reaches its maximum value $2J_2$ and the bands are completely flat.
Fig.~\ref{fig_1}(f) shows the finite $\delta$ case, where all four bands have to be taken into account (same parameters as for Fig.~\ref{fig_1}(e)). For any $\delta$, the gap is again closing at $k_x=\pm\pi/d$ for $\beta=\pi/3$ and at $k_x=0$ for $\beta=2\pi/3$. The maximal gap value is still at $\beta=\pi/2$, but it is smaller than $2J_2$. There is no symmetry within the doublet and the edge states are not at the center of the gap. All these features are summarized in Fig.~\ref{fig_1}(g) showing the energy spectrum versus $\beta$ and the corresponding Zak phase. A trivial gap (zero Zak phase) is present from $\beta=0$ to $\beta=\pi/3$ where the gap closes, then immediately reopens as a topological gap persisting for $\beta \in [\pi/3;2\pi/3]$  (the edge state energy is shown in red). The gap closes at $\beta=2\pi/3$ becoming topologically trivial again. For $\delta < J_1 \cos \beta$, the bands are overlapping (not crossing). There is no real gap anymore, and therefore no protected edge states. 

The equivalence between Hamiltonian~\eqref{reduced_SSH_Hamiltonian} and TFIM spectrum can be exploited to construct and investigate a photonic analog profiting from high accessibility to the spectrum and wavefunction in photonic systems. The TFIM Hamiltonian reads:
\begin{equation}
\label{TFIM_Hamiltonian}
\hat{H}_{TFIM} = -J' \sum_i \sigma_i^z \sigma_{i+1}^z - h \sum_i \sigma_i^x,
\end{equation}
where $J'$ is coupling term, $h$ is a transverse magnetic field and $i$ is the site index. After applying the Jordan-Wigner transformation, it is easy to construct the mapping between Hamiltonians~\eqref{reduced_SSH_Hamiltonian} and \eqref{TFIM_Hamiltonian}: $J'=J_2 \cos \beta_2, h=J_1 \cos \beta_1$. Due to RDSOC, both ferro- and antiferromagnetic configurations are achievable (Fig.~\ref{fig_1}(h)), which is not the case in TFIM analog based on usual SSH. The access to both positive and negative couplings also allows to construct photonic simulators based on Ising Hamiltonian, which is a rapidly developing topic in the last decades~\cite{friedenauer2008simulating,britton2012engineered,marandi2014network,inagaki2016coherent,berloff2017realizing}. This LC-based simulator is compact, can operate at ambient conditions and, in principle, the couplings are reconfigurable by in-plane rotation of LC molecules.

Next, we simulate a realistic implementation of our proposal based on the patterning of the Distributed Bragg Reflector~\cite{scafirimuto2021tunable} of a LC cavity. The 2D potential $U(x,y)$ corresponding to realistic experimental values is shown on Fig.~\ref{fig_2}(a) with the RDSOC being oriented along the strong link of the zigzag chain. We perform simulation beyond the tight-binding approximation by solving the stationary Schrödinger equation with $\hat{H}_{2D} = \hat{H}_{LC} + U$ in 2D (parameters in~\cite{parameters}).

We first consider zero RDSOC realized when polarization bands of same parity are in resonance. Fig.~\ref{fig_2}(c,e) shows the eigenenergies in the real and reciprocal spaces. It shows a clear band gap of 1 meV, but no edge state. Fig.~\ref{fig_2}(d,f) shows the case with non-zero RDSOC ($\alpha=1.62$~meV$\cdot\mu$m) ($N$,$N+1$ case). A mode strongly localized on the edge (also see Fig.~\ref{fig_2}(b)) appears within the gap. The asymmetry between the lower and upper bands is enhanced when going beyond the tight-binding model. However, the lowest band is flattened around $\beta=\pi/2$, as in the tight-binding model (Fig.~\ref{fig_1}(e,f,g)). These simulations clearly demonstrate the possibility to switch between topologically trivial and nontrivial system states by applying voltage.

Finally, we investigate the structure of the polarization and density of the edge state in tight-binding model for the same parameters as for Fig.~\ref{fig_1}(g). Fig.~\ref{fig_3}(a-d) shows the in-plane Stokes vectors for each site \cite{suppl},  and Fig.~\ref{fig_3}(e-h) the intensity distribution.

The circular polarization degree of edge modes is zero because they are localized states with equal contributions of counter-propagating wave vectors.
A clear period doubling effect is observed. To study it, we separate all unit cells into odd and even (Fig.~\ref{fig_3}(i)). For small $\beta$ there is no difference between odd and even unit cells, as one can see in Fig.~\ref{fig_3}(d,h). For $\beta \approx \pi/2$, we see a pronounced difference between odd and even unit cells (Fig.~\ref{fig_3}(c,g)). While all B sites are equivalent (red and green arrows), odd and even A sites (blue and yellow arrows) rotate differently. The density behavior also shows four distinct decay rates. For $\beta = \pi/2$ (Fig.~\ref{fig_3}(b,f)), we have a specific edge state divided into orthogonally-polarized dimers; a single site terminates the chain. The density of even B lattice sites drop almost to zero (thus not visible in Fig.~\ref{fig_3}(f)). 
For even larger $\beta$, the period doubling  disappears. It arises from a superposition of three components of the edge state wavefunction in reciprocal space for $\beta\approx \pi/2$ (Fig.~\ref{fig_3}(j) and details in \cite{suppl}). This occurs because the bulk energy bands exhibit almost equal gaps at the center and edges of BZ (red line in Fig.~\ref{fig_1}(f)).

To conclude, we have shown that RDSOC acts as a synthetic gauge field controlling the magnitude and sign of the tunelling coefficients in a 1D chain. RDSOC controls the chain topology and allows to reach all energy spectra of the different phases of the TFIM. We have proposed a realistic implementation using patterned LC microcavities. Interesting perspectives are related to the realization of 2D lattices, the inclusion of non-Hermitian effects \cite{su2021zzz,krol2021annihilation} and photonic nonlinearities \cite{pernet2022gap}. Also, our proposal can be realized in other platforms such as electronic and fermionic atomic systems with a complete mapping to the TFIM and Kitaev chain Hamlitonians offering interesting perspectives for quantum simulations. 

\begin{acknowledgments}
This work was supported by the European Union Horizon 2020 program, through a Future and Emerging Technologies (FET) Open research and innovation action under Grant Agreement No.~964770 (TopoLight). BP acknowlegde National Science Centre, Poland grant No. 2017/27/B/ST3/00271. 
\end{acknowledgments}


\bibliography{references}

\end{document}